\documentclass{svjour3}

\smartqed

\usepackage{graphicx}
\usepackage{epstopdf}
\usepackage{subfigure}
\usepackage{lmodern}
\usepackage{booktabs}
\usepackage[colorlinks,citecolor=blue,linkcolor=blue,hyperfootnotes=true]{hyperref}
\usepackage[square,sort,comma,numbers]{natbib}
\usepackage{amsmath}
\usepackage{marvosym}

\journalname{Experimental Astronomy}

\begin{document}

\title{On-ground calibrations of the GRID-02 gamma-ray detector}

\author{Huaizhong Gao         \and
            Dongxin Yang            \and
            Jiaxing Wen              \and
            Xutao Zheng             \and
            Ming Zeng                \and
            Jirong Cang             \and
            Weihe Zeng              \and
            Xiaofan Pan              \and
            Qimin Zhou              \and
            Yihui Liu                  \and
            Hua Feng                 \and
            Binbin Zhang            \and
            Zhi Zeng                  \and
            Yang Tian                  \and
            GRID Collaboration
}

\institute{H. Gao \and D. Yang \and J. Wen \and X. Zheng \and W. Zeng \and M. Zeng (\Letter) \and H. Feng \and J. Cang (\Letter) \and Z. Zeng \and Y. Tian \at
              Key Laboratory of Particle and Radiation Imaging (Tsinghua University), Ministry of Education, \\
              Beijing 100084, China \\
              \email{zengming@tsinghua.edu.cn, cangjr14@tsinghua.org.cn}
           \and
              H. Gao \and D. Yang \and J. Wen \and X. Zheng \and W. Zeng \and X. Pan \and Q. Zhou \and Y. Liu \and M. Zeng \and H. Feng \and J. Cang \and Z. Zeng \and Y. Tian \at
              Department of Engineering Physics, Tsinghua University, \\
              Beijing 100084, China \\
            \and
              H. Feng \and J. Cang \at
              Department of Astronomy, Tsinghua University, \\
              Beijing 100084, China \\
            \and
              B. Zhang \at
              Key Laboratory of Modern Astronomy and Astrophysics (Nanjing University), Ministry of Education, \\
              Jiangsu 210093, Nanjing, China
            \and
              B. Zhang \at
              School of Astronomy and Space Science, Nanjing University, \\ Jiangsu 210093, Nanjing, China
            \and
              J. Wen \at
              Science and Technology on Plasma Physics Laboratory, Laser Fusion Research Center, CAEP, \\
              Mianyang 621900, Sichuan, China
}

\date{Received: date / Accepted: date}

\maketitle

\begin{abstract}
  The Gamma-Ray Integrated Detectors (GRID) are a space project to monitor the transient gamma-ray sky in the multi-messenger astronomy era using multiple detectors on-board CubeSats. The second GRID detector, GRID-02, was launched in 2020. The performance of the detector, including the energy response, effective area, angular response, and temperature-bias dependence, is calibrated in the laboratory and presented here. These measurements are compared with particle tracing simulations and validate the Geant4 model that will be used for generating detector responses.
\end{abstract}

\section*{Declarations}
\label{sec:declaration}

\subsection*{Funding}
\label{subsec:funding}
This work is supported by Tsinghua University Initiative Scientific Research Program.

\subsection*{Conflicts of interest}
\label{subsec:conflict_interest}
The authors have no relevant financial or non-financial interests to disclose.

\subsection*{Availability of data and material}
\label{subsec:availability_of_data}
The measured and analyzed data of this work \textcolor{red}{are} available from the corresponding author upon reasonable request.

\subsection*{Code availability}
\label{subsec:code_availability}
The code used for this work is custom made and available from the corresponding author upon reasonable request.

\subsection*{Authors' contributions}
\label{subsec:authors_contributions}
Conceptualization: Ming Zeng, Jirong Cang, Hua Feng, Jiaxing Wen, Donxin Yang; Methodology: Ming Zeng, Hua Feng, Jirong Cang, Dongxin Yang, Huaizhong Gao; Funding acquisition: Ming Zeng, Hua Feng, Zhi Zeng, Yang Tian; Formal analysis and investigation: Huaizhong Gao, Dongxin Yang, Jiaxing Wen, Jirong Cang, Weihe Zeng, Xiaofan Pan, Qimin Zhou, Yihui Liu; Writing - original draft preparation: Huaizhong Gao, Dongxin Yang, Ming Zeng, Jirong Cang; Writing - review and editing: Ming Zeng, Hua Feng, Jirong Cang, Binbin Zhang, Dongxin Yang, Huaizhong Gao; Resources: Ming Zeng, Hua Feng, Jirong Cang, Zhi Zeng; Supervision: Ming Zeng, Hua Feng, Jirong Cang, Binbin Zhang, Zhi Zeng, Yang Tian

\section{Introduction}
\label{sec:intro}

The Gamma-Ray Integrated Detectors (GRID) are a space project conducted by students to detect transient gamma-ray events in the energy range from 10~keV to 2~MeV. GRID is a network of small detectors deployed in low Earth orbits using CubeSats with an all-sky coverage. The primary science targets of GRID include the gamma-ray bursts (GRBs) and soft gamma-ray repeaters (SGRs). Other transients like the terrestrial gamma-ray flashes (TGFs) and solar flares can also be observed and studied with GRID. The first GRID detector prototype, GRID-01, was launched on October 29, 2018 \cite{wen_grid_2019}. GRID-01 validates the design of the detector. GRID-02 is an improved version expected to collect useful science data.

\begin{figure}
  \centering
  \includegraphics[scale=0.4]{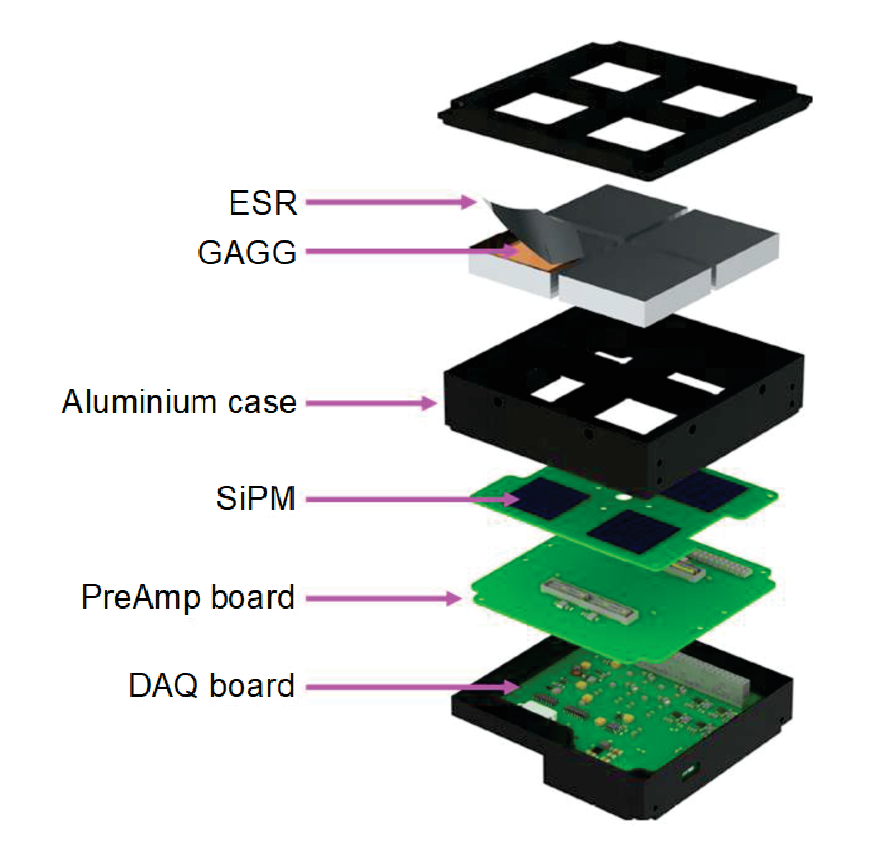}
  \caption{Structure of \textcolor{red}{the} GRID detector, adopted from Wen et al.~\cite{wen_grid_2019}}.
  \label{fig:structure}
\end{figure}

The GRID-02 detector was assembled in 2019. A standard GRID detector design was adopted, shown in Fig.~\ref{fig:structure}. The gamma-ray detection unit includes an array of silicon photomultipliers (SiPM) coupled with a piece of Ce-doped ${\rm Gd}_{3}{\rm (Al, Ga)}_{5}{\rm O}_{12}$ (GAGG) scintillation crystals. Four such units are mounted in $2 \times 2$ pattern, on top of a preamplifier (PreAmp) board and a data acquisition (DAQ) board. Each detector unit is connected with an independent readout. A proportion integral differential (PID) control for the SiPM operating bias voltage is introduced to stabilize the bias against the increase in the \textcolor{red}{leakage} current caused by cosmic radiation. The general properties of GRID-02 are listed in Table~\ref{tab:parameters}, and a more detailed description of the instrument can be found in \cite{wen_compact_2021}. Launched in November 2020, GRID-02 soon became fully operational and has been accumulating science data since then.

As the detector and surrounding materials are not spherically symmetric, transient sources at different incident directions will result in different fluxes and energy spectra. Thus, in order to model and reconstruct the incident energy spectrum, \textcolor{red}{the detector responses as a function of incident directions are required} \cite{wen_grid_2019,bissaldi_ground-based_2009}. This is possible with particle tracing modelling of the detector, which should be validated with laboratory tests. In this paper, we present the calibration campaigns, data analysis, and results.

\begin{table}[htb]
  \centering
  \caption{Basic properties of the GRID-02 detector.}
  \label{tab:parameters}
  \begin{tabular}{cc}
    \toprule\noalign{\smallskip}
    Detector size & 0.5~U ($9.4 \times 9.4 \times 5~{\rm cm}^{3}$)  \\
    \noalign{\smallskip}\hline\noalign{\smallskip}
    Weight & 780~g \\
    \noalign{\smallskip}\hline\noalign{\smallskip}
    Power & $\leq 3~{\rm W}$ \\
    \noalign{\smallskip}\hline\noalign{\smallskip}
    Detection area & $58~{\rm cm}^{2}$ \\
    \noalign{\smallskip}\hline\noalign{\smallskip}
    FOV & \textcolor{red}{$2~\pi$} \\
    \noalign{\smallskip}\hline\noalign{\smallskip}
    Energy range & 10~keV--2~MeV \\
    \noalign{\smallskip}\hline\noalign{\smallskip}
    Dead time & $20~\mu{\rm s}$ \\
    \noalign{\smallskip}\bottomrule
  \end{tabular}
\end{table}

\section{The calibration setup}
\label{sec:setup}

\subsection{Tests with radioactive sources}
\label{sec:setup_radioactive_source}

\begin{figure}
  \centering
  \includegraphics[scale=0.6]{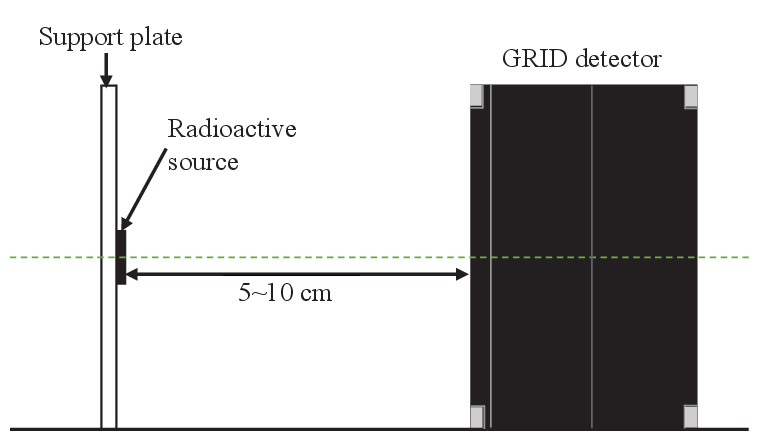}
  \caption{Setup of the energy response calibration. The radioactive source is mounted on a supporting plate, at a distance of about 5--10~cm from the GRID detector. The green dashed line marks the axis of the detector.}
  \label{fig:source_setup}
\end{figure}

Calibration of the energy response is conducted in the energy range from 60~keV to 1.3~MeV using five radioactive sources (Table~\ref{tab:source_properties}) in the laboratory. The setup of the calibration experiment is shown in Fig.~\ref{fig:source_setup}. A weak radioactive source was mounted on a support plate and placed on the axis of the detector, at a distance of approximately 5--10~cm. The background spectra without the sources were measured and subtracted. The activity of the sources was not determined, and therefore we were unable to calculate the detector's effective area at these energies.

\begin{table}[htb]
  \centering
  \caption{Radioactive sources used for energy response calibration.}
  \label{tab:source_properties}
  \begin{tabular}{cc}
    \toprule\noalign{\smallskip}
    Nuclide & Gamma-ray energy (keV)  \\
    \noalign{\smallskip}\midrule\noalign{\smallskip}
    $^{241}{\rm Am}$ & 59.5 \\
    $^{212}{\rm Pb}$ (within $^{228}{\rm Th}$ source) & 238.6 \\
    $^{22}{\rm Na}$ & 511.0 \\
    $^{137}{\rm Cs}$ & 661.7 \\
    $^{60}{\rm Co}$ & 1332.5 \\
    \noalign{\smallskip}\bottomrule
  \end{tabular}
\end{table}

\subsection{Tests with the X-ray beam}
\label{sec:setup_xray}

\begin{figure}
  \centering
  \includegraphics[scale=0.6]{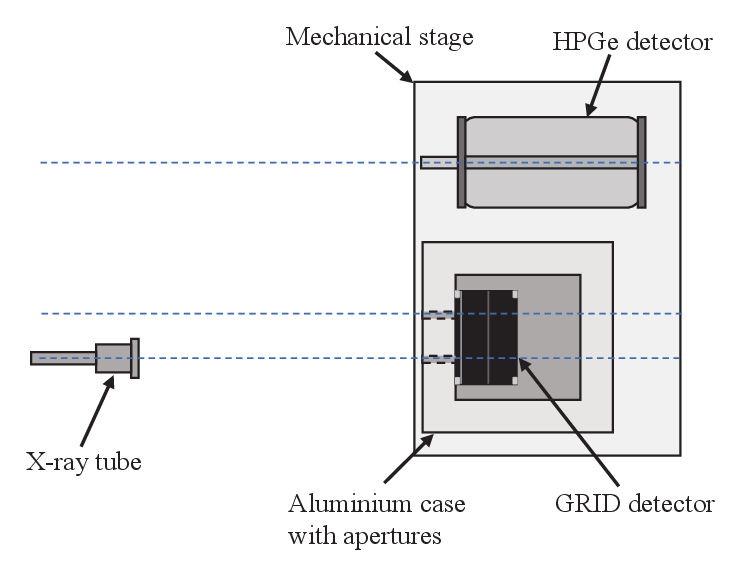}
  \caption{Setup of the X-ray beam test at NIM, viewed from top. The GRID detector is placed within the aluminium case with apertures at the front. A calibrated HPGe detector is mounted on a mechanical stage. The blue dashed lines mark the center positions of the four GAGG crystals of the GRID detector and the axis of the HPGe detector.}
  \label{fig:setup_xray}
\end{figure}

The beam test is performed with the Hard X-ray Calibration Facility (HXCF) at the National Institute of Metrology (NIM) in Beijing. The experimental setup is shown in Fig.~\ref{fig:setup_xray}. The beam size is approximately 6~mm. The GRID detector is mounted in an aluminium case, on top of which there are four apertures with a diameter of 10~mm at positions corresponding to the centres of the four GAGG crystals. \textcolor{red}{Extra steps have been taken to ensure that the photon energy is fully deposited in the sensitive volume of the detector before each measurement.} The four detector units are exposed to the X-ray beam one after another. After that, a calibrated high-purity germanium (HPGe) detector serving as a standard detector is moved to the beam to acquire a \textcolor{red}{reference spectrum}. The effective area of the GRID detector is calculated based on the \textcolor{red}{reference spectrum}. This experiment covers an energy range from 13.4~keV to 120~keV.

\subsection{Calibration of the detector gain}
\label{sec:setup_tempbias}

\begin{figure}
  \centering
  \includegraphics[scale=0.6]{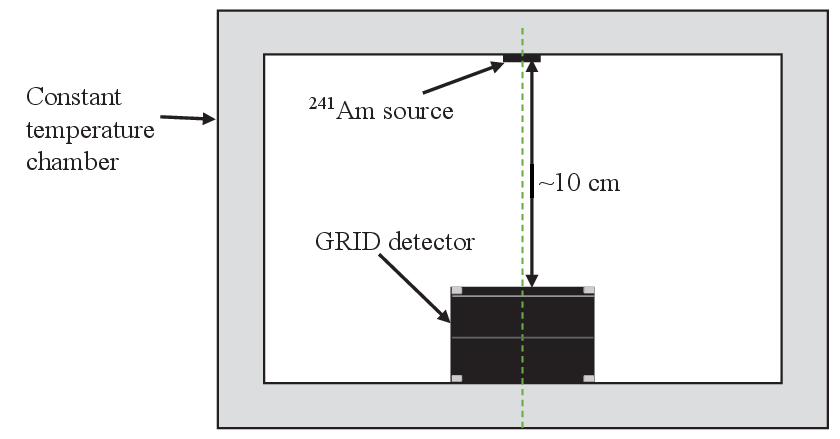}
  \caption{Setup of the test at different temperatures and bias. Both the $^{241}{\rm Am}$ source and GRID detector sit in a constant temperature chamber. The dashed line marks the axis of the GRID detector.}
  \label{fig:setup_tempbias}
\end{figure}

Calibration of the temperature/bias dependence of the detector gain is conducted in a constant temperature chamber. The experimental setup is shown in Fig.~\ref{fig:setup_tempbias}. An $^{241}{\rm Am}$ source is mounted above the detector axis. Measurements are performed at different temperatures and SiPM bias voltages. The \textcolor{red}{measuerd temperature of the SiPM} ranges from 0~$^{\circ}{\rm C}$ to 30~$^{\circ}{\rm C}$ and the SiPM bias voltage ranges from 27.0~V to 29.0~V. The data are acquired for each detector unit. The detector gain as a function of the temperature and bias is obtained. \textcolor{red}{Note that all the temperature values to be mentioned in the following sections refer to the measured temperature by the on-board temperature sensors, which are close to each SiPM units. Since the gain correction for the subsequent in-orbit data is also performed using the measured SiPM temperature and bias voltage values, the calibration result for the detector gain can be directly used for future data correction.}

\subsection{Calibration of the angular responses}
\label{sec:setup_angle}

\begin{figure}
  \centering
  \subfigure{\includegraphics[height=6cm]{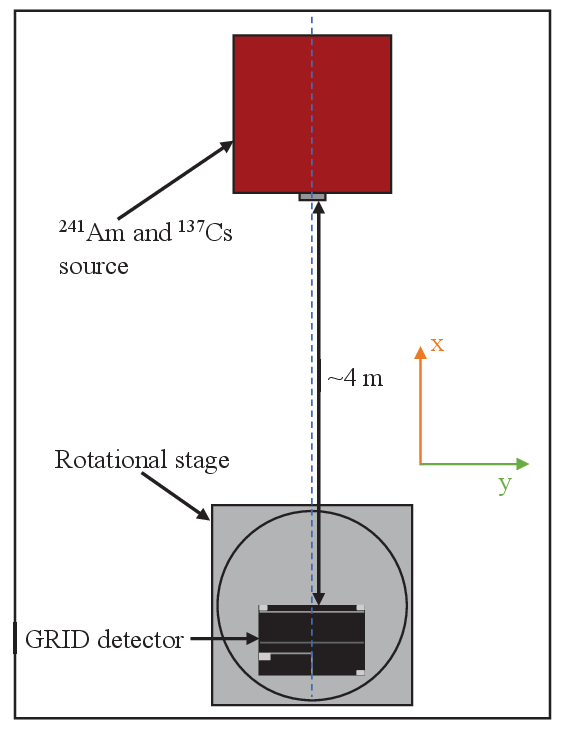}}
  \subfigure{\includegraphics[height=6cm]{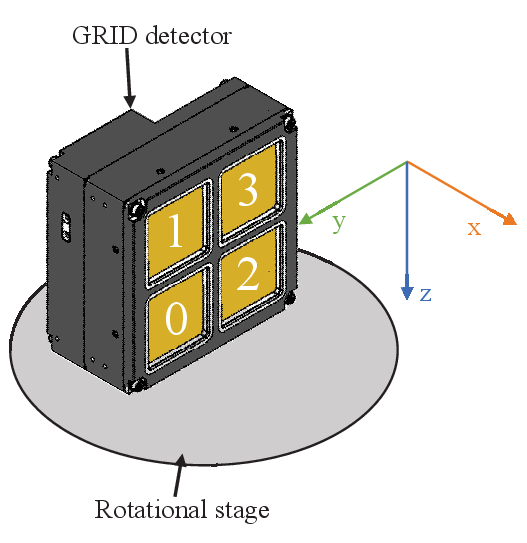}}
  \caption{Setup of the angular responses experiment.}
  \label{fig:setup_angle}
\end{figure}

The effective area of the detector is a function of the incident direction of gamma-rays, due to different absorbing materials and projected areas of the sensor. This needs be calibrated and is referred to as the angular responses of the detector. The setup is shown in Fig.~\ref{fig:setup_angle}. The detector is mounted on a rotational stage. The source used in the experiment is a combination of two high-activity radioactive sources, $^{241}{\rm Am}$ (100~mCi) and $^{137}{\rm Cs}$ (10m~Ci), at a distance of approximately 4 m from the detector, such that the whole detector can be irradiated uniformly. Measurements are obtained at an angle cadence of 15$^{\circ}$ from 0$^{\circ}$ to 360$^{\circ}$. Simulations using Geant4 with the same setup are also performed. 

\section{Data analysis and results}
\label{sec:results}

\begin{figure}
  \centering
  \subfigure{\includegraphics[scale=0.37]{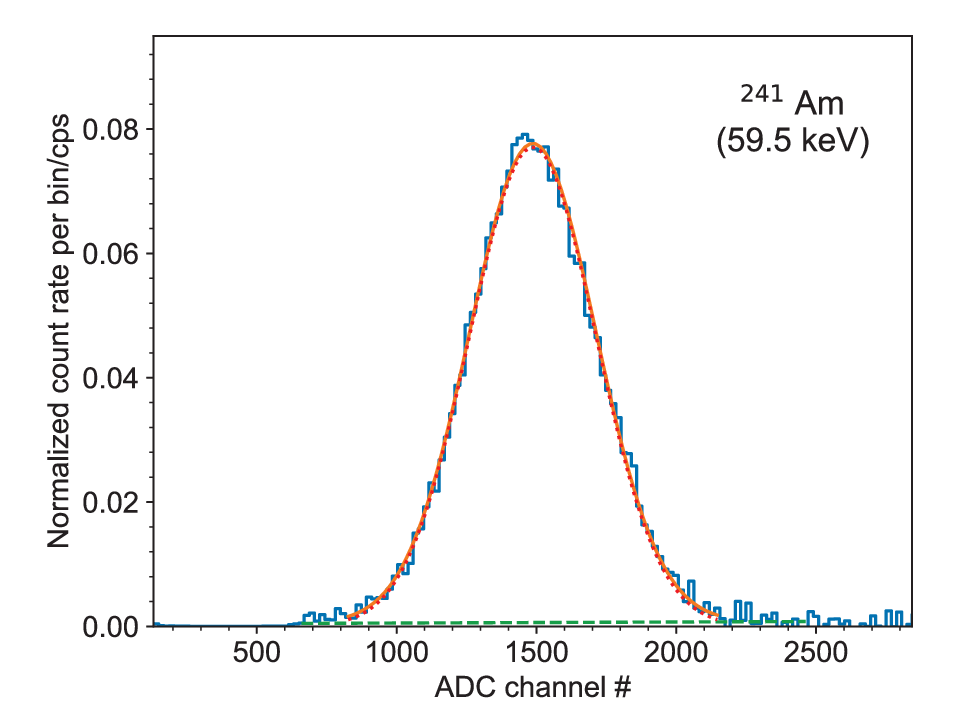}}
  \subfigure{\includegraphics[scale=0.37]{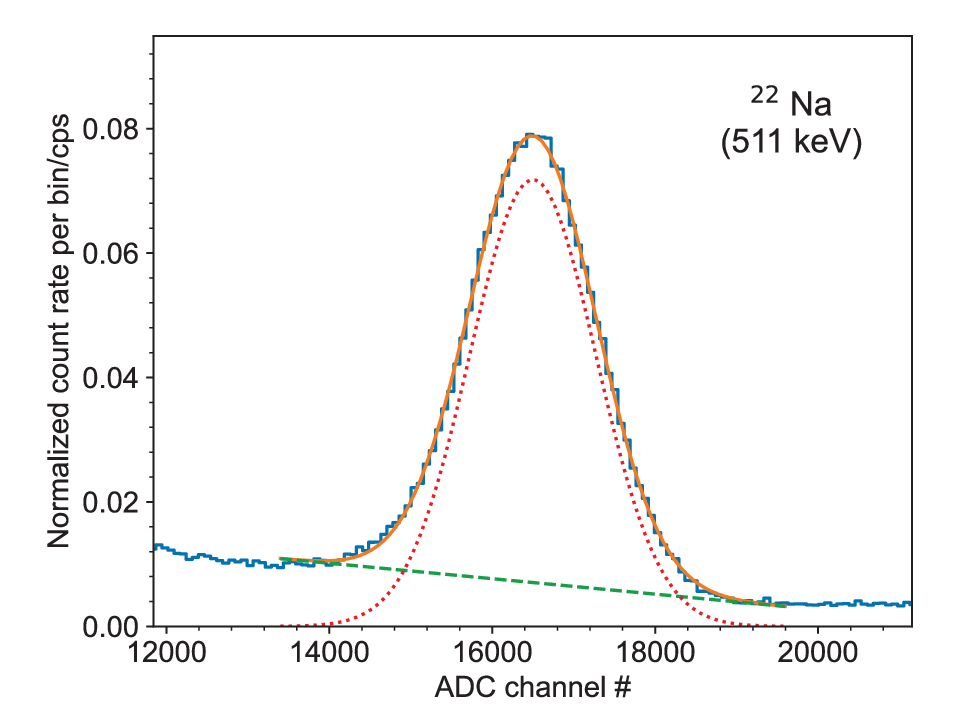}}
  \caption{Example energy spectra with model decomposition. Each spectrum is fitted with a Gaussian plus a linear component to account for the local background.}
  \label{fig:fit_demo}
\end{figure}

The spectrum due to monochromatic X/$\gamma$-rays is fitted with a Gaussian to find the energy peak. \textcolor{red}{Besides, the contributions from the further non-photopeak background are modeled as a linear around the peak.} Two spectra with best-fit results are demonstrated in Fig.~\ref{fig:fit_demo}. In some cases, \textcolor{red}{e.g. the 1173.2~keV line of $^{60}{\rm Co}$}, the Compton component from high energy photons or the readout noise along with low energy peaks are significant in the residual; they are considered and removed if needed \cite{bissaldi_ground-based_2009}.  

\subsection{Gain vs.\ temperature and bias}
\label{subsec:tempbias}

\textcolor{red}{The gain of the GRID-02 detector has a dependence on the operating temperature and the SiPM bias voltage. Therefore, the dependence of the detector gain on temperature and the SiPM bias must be first determined. Then, in order to eliminate the gain difference caused by the variation of the operating temperature, the gain correction is performed for the rest of the calibrations.}

\textcolor{red}{The light yield of GAGG has a strong dependence on temperature \cite{nakajima_temperature_2019,yoneyama_evaluation_2018}}, and can be described with the following function 
\begin{equation}
  {\rm LY}_{\rm GAGG} \approx a_{\rm T} \cdot T^{2} + b_{\rm T} \cdot T + c_{\rm T} \, ,
\label{eq:light_yield}
\end{equation}
where $a_{\rm T}$, $b_{\rm T}$, and $c_{\rm T}$ are coefficients from Taylor expansion.

The amplitude of the SiPM output is proportional to the photon detection efficiency (PDE) and gain, which can both be determined by the overvoltage of the SiPM
\begin{equation}
  V_{\rm OV} = V_{\rm b} - V_{\rm BD} \label{eq:vov} \, ,
\end{equation}
where $V_{\rm b}$ is the SiPM bias, and $V_{\rm BD}$ is the breakdown voltage. The breakdown voltage is related to the operating temperature following
\begin{equation}
  V_{\rm BD} = k \cdot T + V_{\rm BD0} \label{eq:vbd} \, .
\end{equation}

The PDE of the SiPM can be expressed as
\begin{equation}
  {\rm PDE} = {\rm PDE}_{\max} \cdot (1 - e^{-V_{\rm OV}/V_{\rm p}}) \, , \label{eq:pde}
\end{equation}
where ${\rm PDE}_{\max}$ is the maximum PDE, and $V_{\rm p}$ is the wavelength-dependent growth constant \cite{zappala_study_2016,otte_characterization_2017}. The gain of the SiPM is proportional to its overvoltage \cite{otte_characterization_2017}, %$G_{\rm SiPM} \propto V_{\rm OV}$
\begin{equation}
  G_{\rm SiPM} \propto V_{\rm OV} \, , \label{eq:gain_sipm}
\end{equation}

The detector gain can be expressed as the product of the light yield, PDE, the SiPM gain, and the electronics gain,
\begin{equation}
  G_{\rm det} = {\rm LY}_{\rm GAGG} \cdot {\rm PDE} \cdot G_{\rm SiPM} \cdot G_{\rm elec} \label{eq:gain} \, ,
\end{equation}
and $G_{\rm elec}$ can be regarded as a constant. Combining the above equations, the detector gain can be described as a function of the temperature and SiPM bias,
\begin{align}
  G_{\rm det}(T, V_{\rm b}) &= G_{0} \cdot V_{\rm OV} \cdot (1 - e^{-{V_{\rm OV}}/{V_{\rm p}}}) \cdot (-T^{2} + b_{\rm T} \cdot T + c_{\rm T}) \nonumber \\
   &= G_{0} \cdot (V_{\rm b} - k \cdot T - V_{\rm BD0}) \cdot (1 - e^{-(V_{\rm b} - k \cdot T - V_{\rm BD0})/V_{\rm p}}) \nonumber \\
   &\quad \cdot (-T^{2} + b_{\rm T} \cdot T + c_{\rm T}) \label{eq:gain_theoretical} \, ,
\end{align}
where $G_0$ absorbs the constant terms and represents the gain at the standard temperature and bias voltage. Considering $V_{\rm OV} \ll V_{\rm p}$, the above equation can be simplified as 
\begin{align}
  G_{\rm det}(T, V_{\rm b}) &= G_{0} \cdot V_{\rm OV}^{2} \cdot (-T^{2} + b_{\rm T} \cdot T + c_{\rm T}) \nonumber \\
  &= G_{0} \cdot (V_{\rm b} - k \cdot T - V_{\rm BD0})^{2} \cdot (-T^{2} + b_{\rm T} \cdot T + c_{\rm T}) \label{eq:gain_empirical} \, .
\end{align}

This equation is used to fit the data. We adopt $k = 21.5$~mV/$^{\circ}$C as specified in the SiPM datasheet\footnote{\url{https://www.onsemi.com/pdf/datasheet/microj-series-d.pdf}.}.

\begin{figure}
  \centering
  \includegraphics[scale=0.6]{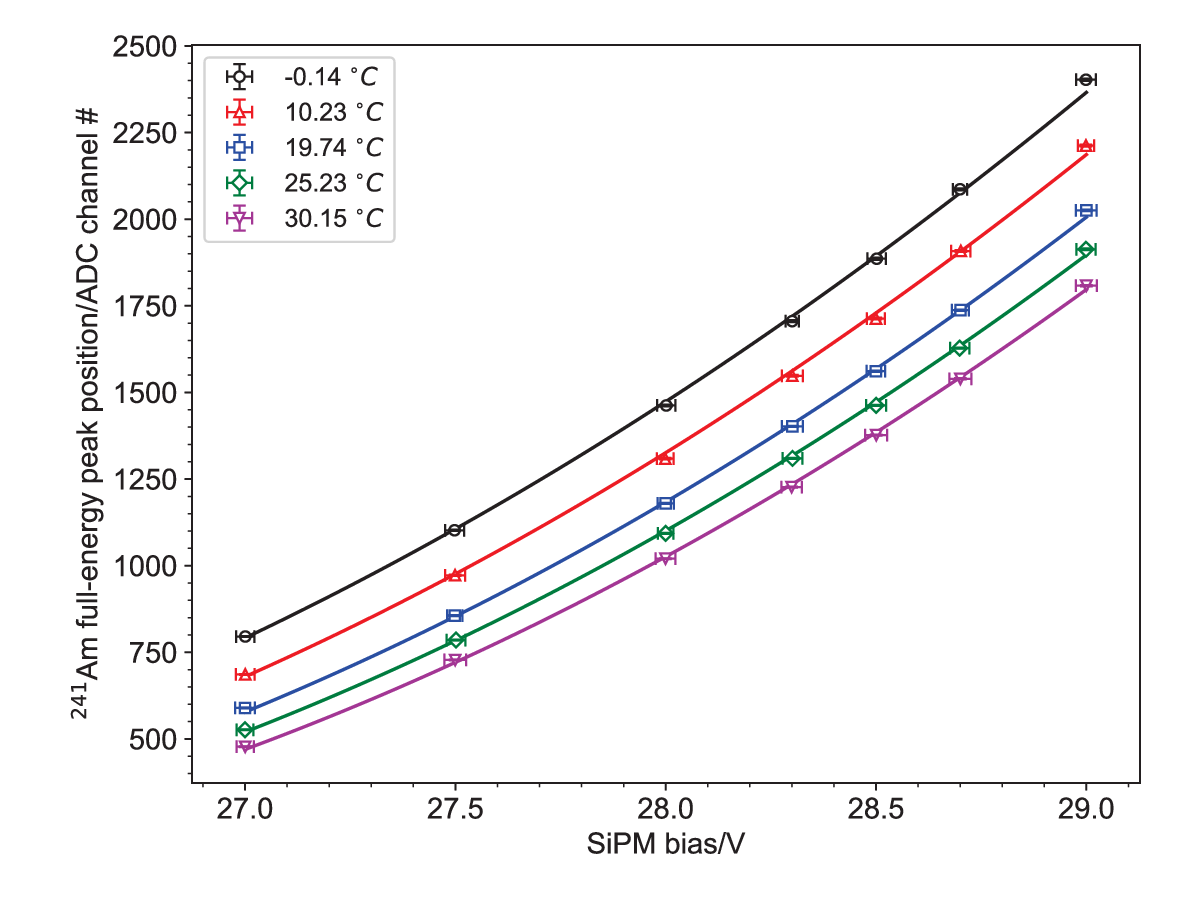}
  \caption{Fit result of the temperature and bias dependence of channel 0. The overall dependence is expressed in the form of the bias dependence (gain versus SiPM bias) at different temperatures.}
  \label{fig:tempbias}
\end{figure}

The detector gain is measured at various temperatures (from about 0 to 30~$^\circ$C) and SiPM biases (27--29~V) using the $^{241}$Am radioactive source. The results are shown in \textcolor{red}{Fig.~\ref{fig:tempbias}} with the best-fit models. The systematic error, which is approximately 0.5~$^{\circ}$C introduced by the temperature sensor, has been considered. The fit residuals are $\leq6\%$ for all the four units across the above parameter space. Around the standard operating bias (28.5~V), the residual is reasonably small ($<3\%$). \textcolor{red}{This allows us to correct the gain based on the measured temperature and bias in the following calibrations and the in-orbit data.}

\textcolor{red}{For the data processing of the calibrations in the following sections, the gain correction is performed using the above-mentioned gain calibration result. By normalizing the detector gain to the value at standard operating temperature (25~$^{\circ}$C) and bias (28.5~V), the difference in the detector gain between each measurement can be eliminated.}

The fit residual is quite large (around 6\%) at some marginal data points, e.g. data points with low biases near 27.0~V. This indicates that Eq.~\ref{eq:gain_empirical} is still not a sufficiently accurate approximation of the actual detector response. \textcolor{red}{We are currently working on improving the fit model and the result will be reported elsewhere.}

\subsection{Spectral calibration}
\label{sec:spectral}

\subsubsection{Channel-energy relation}
\label{sec:channel_energy}

\begin{figure}
  \centering
  \subfigure{\includegraphics[scale=0.37]{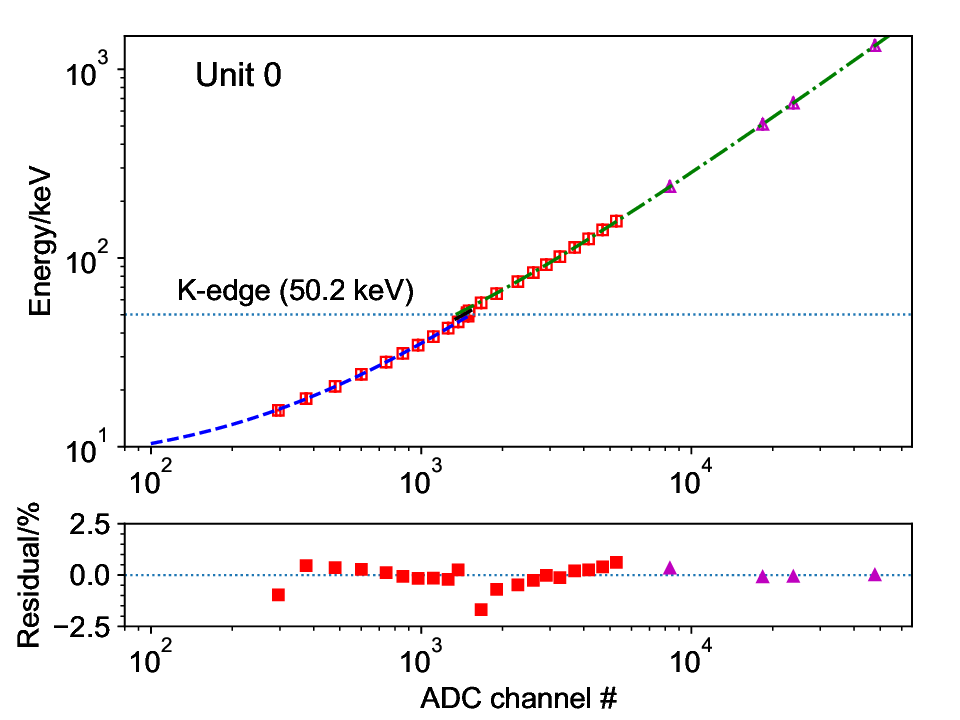}}
  \subfigure{\includegraphics[scale=0.37]{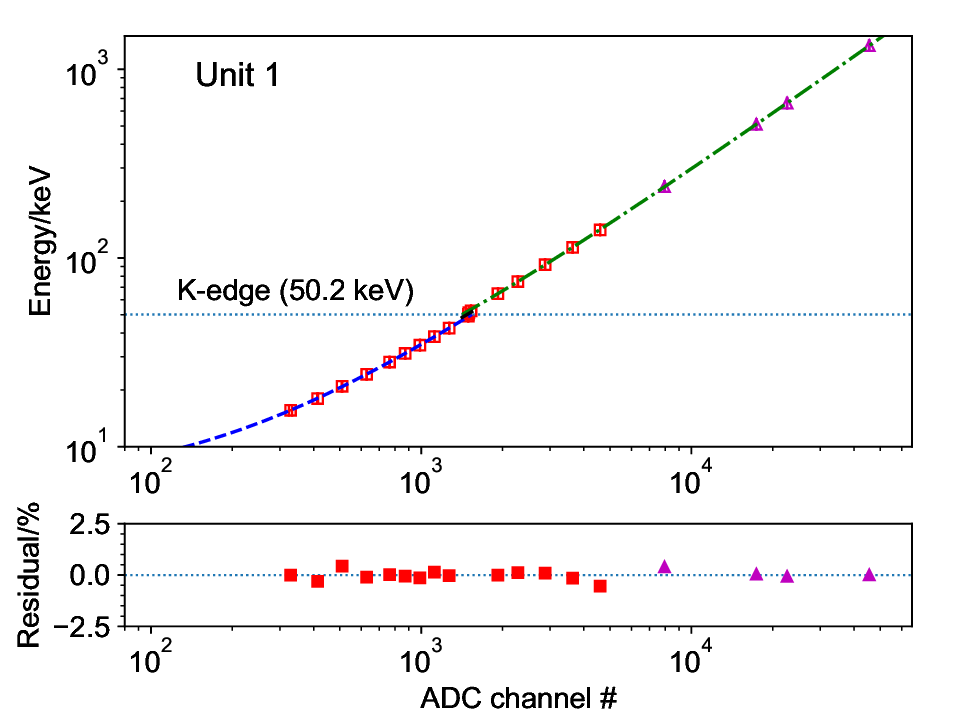}}
  \subfigure{\includegraphics[scale=0.37]{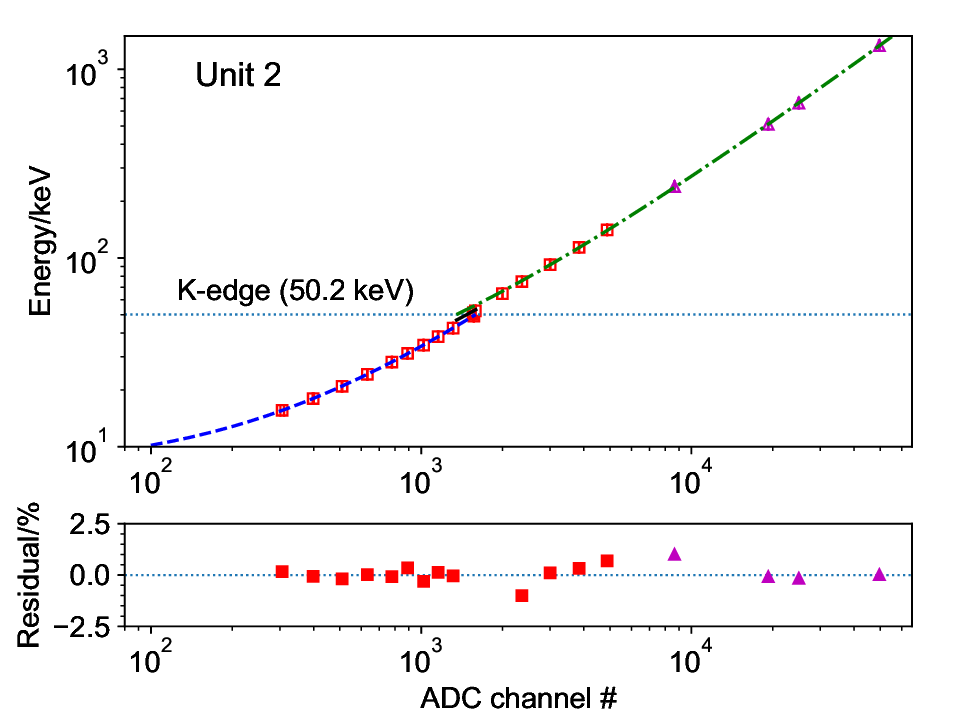}}
  \subfigure{\includegraphics[scale=0.37]{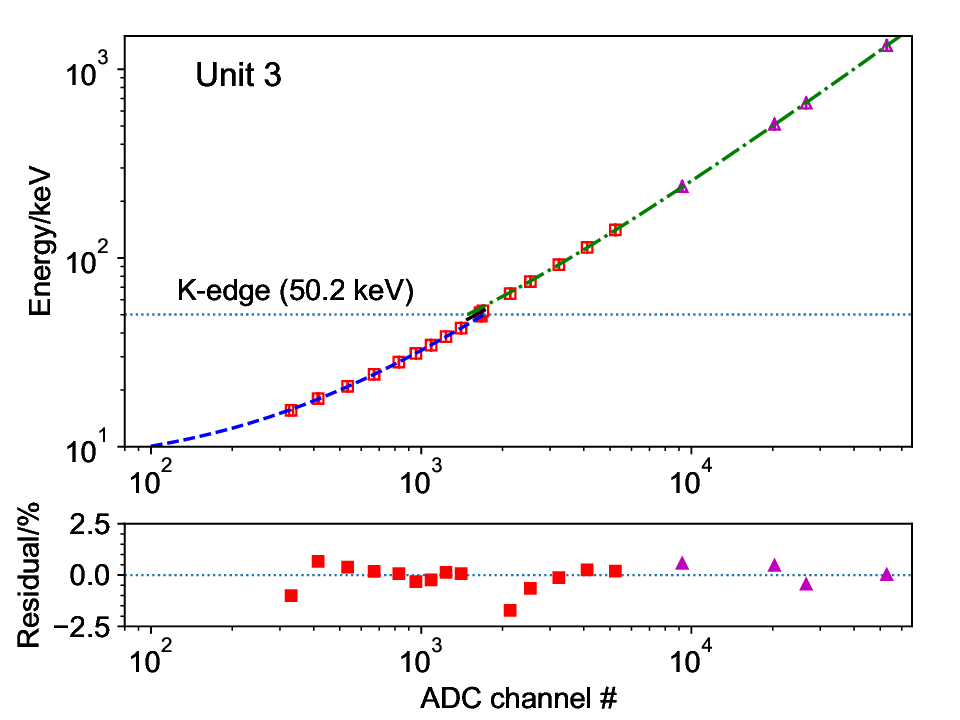}}
  \caption{Channel-energy relations and fit residuals of the four detector units. The squares indicate the data from the X-ray beam, and the triangles indicate measurements with the radioactive sources. The relations in the low and high energy regions are marked with dashed and dotted-dashed lines, respectively. The horizontal dotted blue line indicates the energy of the gadolinium K-edge.}
  \label{fig:channel_energy}
\end{figure}

\begin{figure}[htb]
  \centering
  \includegraphics[scale=0.6]{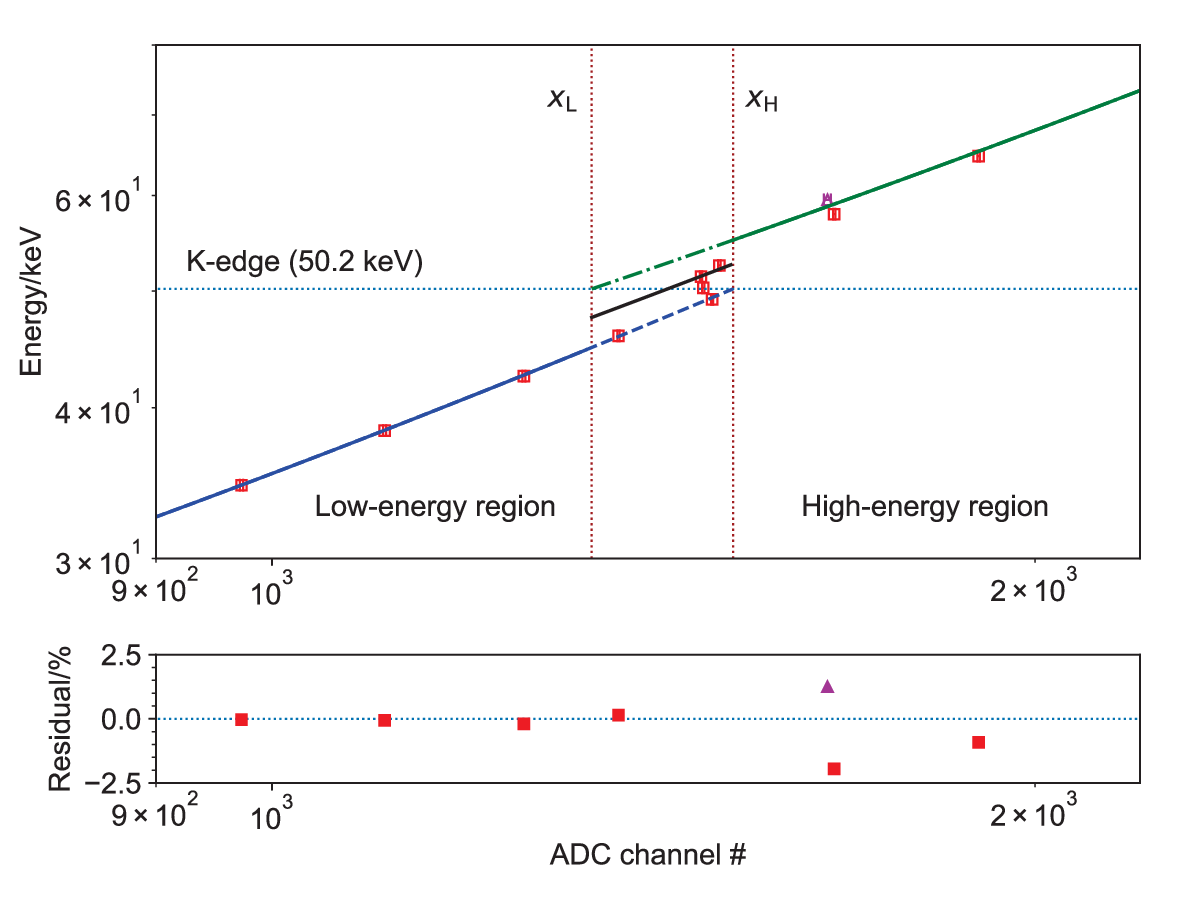}
  \caption{Channel-energy relation near the gadolinium K-edge (50.2~keV). The data are adopted from detector unit 0. The quadratic relation in the low and high energy bands is plotted with blue and green lines, respectively. The solid black line, which is the mean of the low and high energy relations, represents the relation around the K-edge. The horizontal line represents the K-edge energy and the two vertical lines represent the lower and upper bounds of the K-edge band.}
  \label{fig:channel_energy_kedge}
\end{figure}

Because the light yield of GAGG is a function of energy, the channel-energy relation has some degree of nonlinearity, e.g., there is a drop of the light yield near the K-edge of gadolinium (50.2~keV) \textcolor{red}{\cite{ferreira_energy_2004}}. The tests are done using both the X-ray beam and radioactive sources; the former covers an energy range from 13.4--49.0~keV, and the latter covers 55.0~keV--1.332~MeV. As the two types of sources may have different illuminating geometries, there is a discrepancy \textcolor{red}{of} less than 5\% between them. \textcolor{red}{A quadratic polynomial is used to fit the relation, but there is a discrepancy for data below and above the gadolinium K-edge} (see Fig.~\ref{fig:channel_energy}). A narrow band around the K-edge is not used for fitting because the data in this region apparently deviate from the relation extrapolated from either the low or high energy band. In this narrow band, we adopt \textcolor{red}{the} mean of the two relations. This is illustrated in Fig.~\ref{fig:channel_energy_kedge}. To summarize, a three-segment (below, around, and above the gadolinium K-edge) channel-energy relation is derived for each detector unit. 

\subsubsection{Energy resolution}
\label{sec:resolution}

\begin{figure}
  \centering
  \includegraphics[scale=0.6]{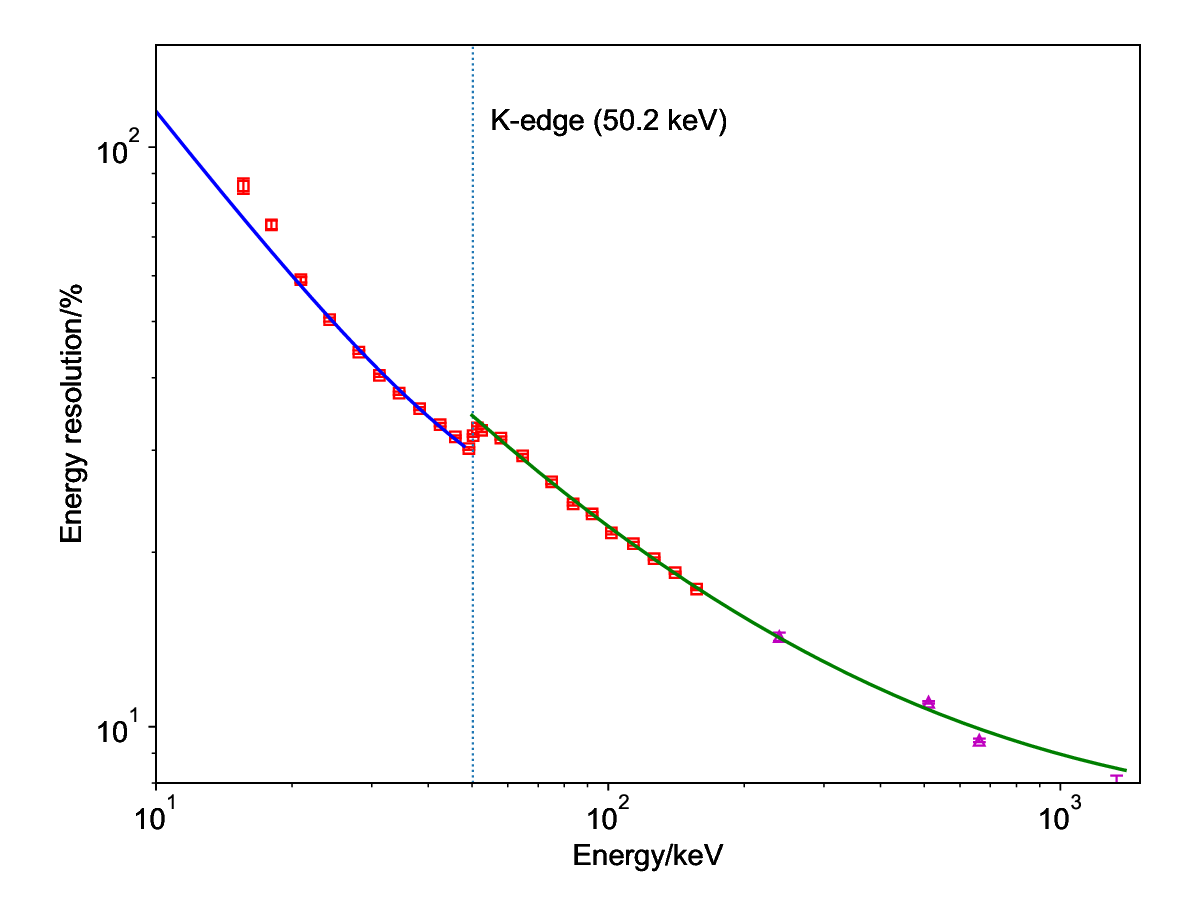}
  \caption{Spectral resolution as function of energy for detector unit 0. The data from the beam test are marked with square markers, while those with the radioactive sources are marked with triangles. The vertical dotted blue line marks the energy of the gadolinium K-edge.}
  \label{fig:resolution}
\end{figure}

The energy resolution in terms of the full width at half maximum (FWHM) is measured at different energies. Its dependence with energy is displayed in Fig.~\ref{fig:resolution} and fitted with an empirical function,
\begin{equation}
  {\rm FWHM} = \sqrt{a_{\rm R} \cdot \left(\frac{E}{\rm keV}\right)^{2} + b_{\rm R} \cdot \left(\frac{E}{\rm keV}\right) + c_{\rm R}} \, ,
\label{eq:fwhm}
\end{equation}
where $a_{\rm R}$, $b_{\rm R}$, and $c_{\rm R}$ are coefficients that represent the intrinsic resolution of GAGG caused by non-ideal transport efficiency of scintillation photons, statistical fluctuations, and electronic noise, respectively \cite{bissaldi_ground-based_2009}. There is a deviation of the resolution data at low energies, causing a discrepancy of less than 15\%. Such deviation can be explained by a possible electronic noise component, which broadens the peak of some low-energy X-ray beams. The overall energy resolution is derived with the fitting results below and above the K-edge, yielding a relatively good resolution of 9.5\% at 661.7~keV.

\subsection{Effective area and angular response}
\label{subsec:effective_area}

\begin{figure}
  \centering
  \includegraphics[scale=0.6]{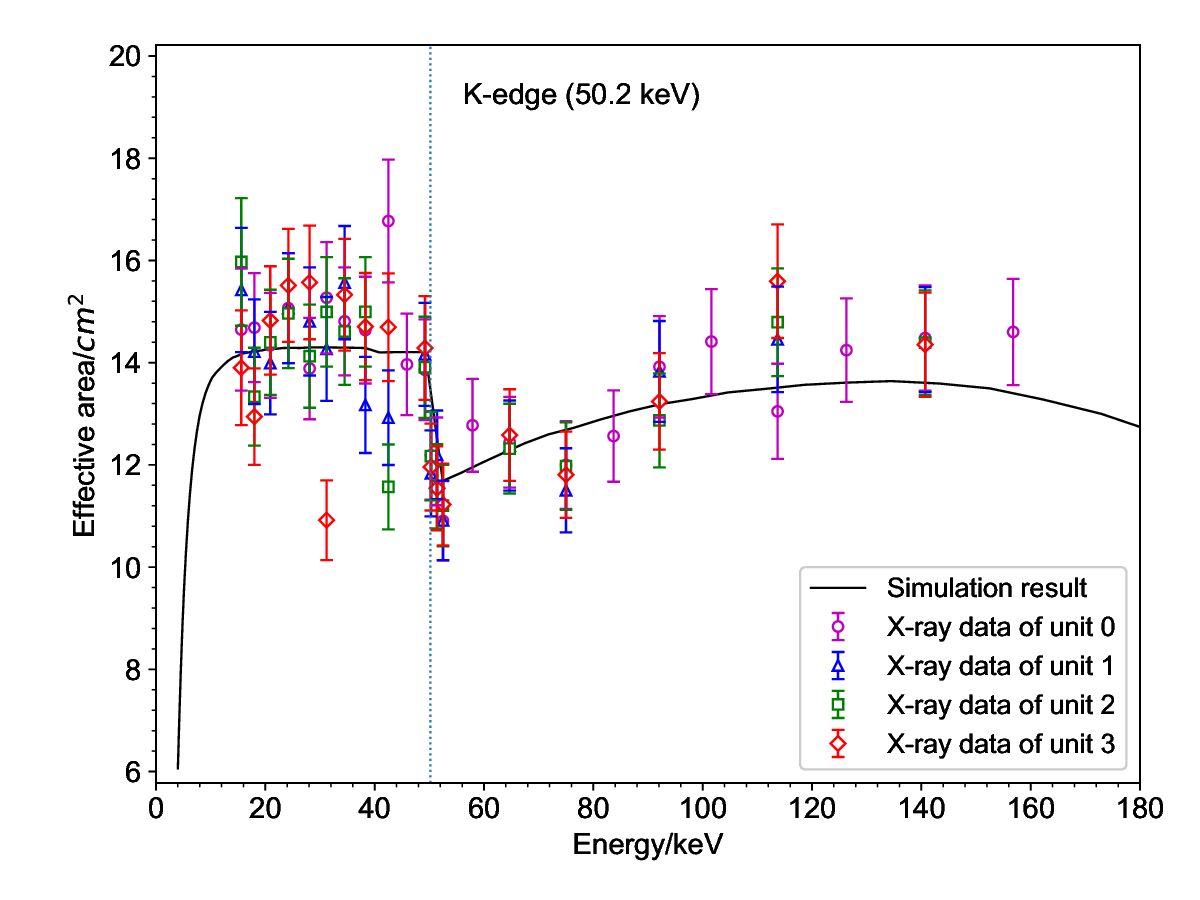}
  \caption{Measured and simulated effective area as a function of energy. A systematic error of around 7.1\% caused by the beam instability is added to the error bar. The vertical line marks the energy of the gadolinium K-edge.}
  \label{fig:effective_area_low_energies}
\end{figure}

The on-axis effective area is measured with the X-ray beam, calibrated against the standard HPGe detector. The results are shown in Fig.~\ref{fig:effective_area_low_energies} in comparison with the simulated effective area using Geant4. We note that the flux of the X-ray beam is unstable and further investigation reveals a systematic error of 5\%. Since the fluctuation of beam intensity is equivalent to the measurement of a data point in time scale, the systematic error of 5\% has been added into the error budget of the data from both detectors. Consequently, a systematic error of around 7.1\% is added to the effective area result. Although the measurements have relatively large uncertainties, the data agree with the simulation result within errors with \textcolor{red}{an average} discrepancy of 5.9\%, especially around the K-edge of gadolinium. \textcolor{red}{For the calibration of existing space gamma-ray observation missions, e.g. Fermi GBM \cite{bissaldi_ground-based_2009}, the measurement error of effective area should be less than 10\%. Therefore, the calibration result of the on-axis effective area is considered to be acceptable, especially for the comparison with Geant4 simulations.} This implies that the simulation can be used to generate the spectral response. 

\begin{figure}
  \centering
  \subfigure{\includegraphics[scale=0.37]{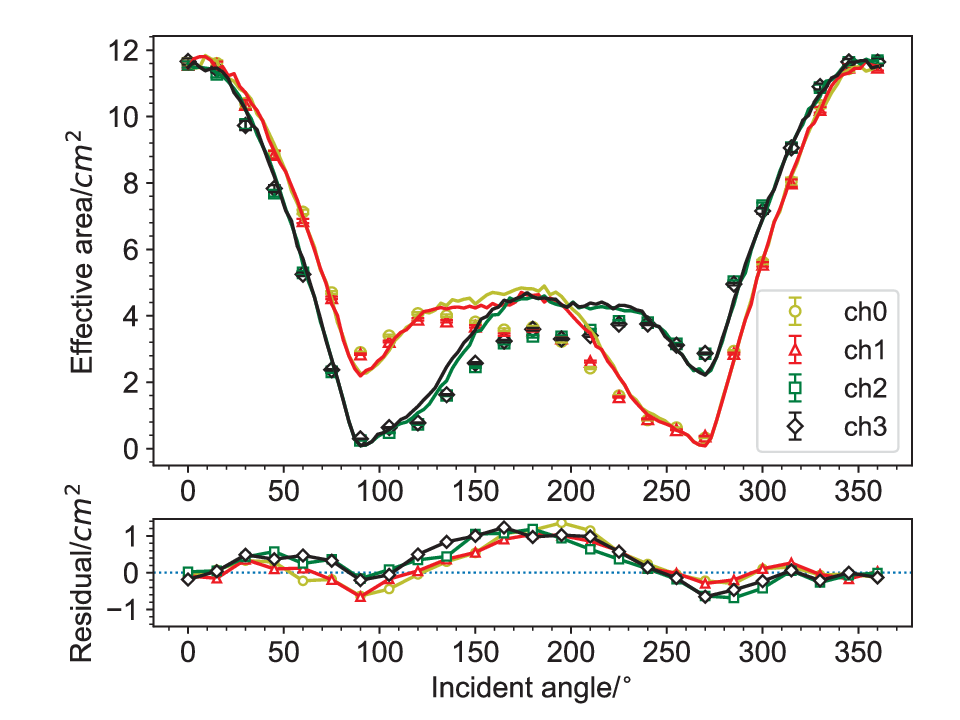}}
  \subfigure{\includegraphics[scale=0.37]{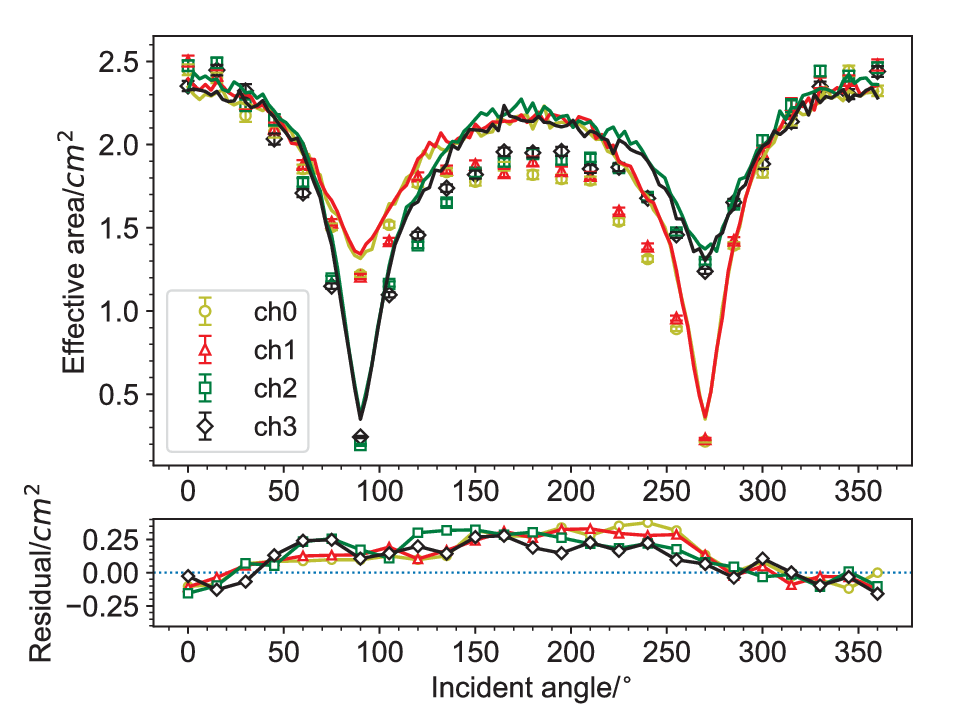}}
  \caption{Measured (points) and simulated (curve) angular responses of the detector, i.e., the effective area with a function of incident angles. The response measured with $^{241}{\rm Am}$ is shown on the left panel, and that measured with $^{137}{\rm Cs}$ is on the right.}
  \label{fig:angular_response}
\end{figure}

The effective area at different incident angles, i.e., its angular response, is measured using radioactive sources as shown in Fig.~\ref{fig:angular_response}. As the experiments are done in the laboratory without a standard detector, the measured angular response is normalized to the simulated curve at angles in the range of 0--90$^\circ$ and 270--360$^\circ$. \textcolor{red}{Due to} two radioactive sources in the setup, the effective area at two energies (59.5 and 661.7~keV) can be derived with each measurement. As one can see, the measured angular response agrees with the simulation at small or large incident angles (close to face-on), with \textcolor{red}{an} average discrepancy of 3.2\% at 59.5~keV and 4.4\% at 661.7~keV. However, for back-incident (near 180$^\circ$) gamma-rays, the measurement deviates from the simulation by a factor far larger than that of the front-incident case. This is because the mass model may be inaccurate due to complicated structures in the detector beneath the GAGG sensor. \textcolor{red}{Since the detector is designed to be gamma-ray sensitive with a $2~\pi$ FOV, the observations will mainly be conducted at front-incident angles, while only the high-energy photons can penetrate the materials at the back of the detector. Therefore, the consistency of measured and simulated angular response is evaluated using the front-incident data.} A dedicated study to improve the Geant4 modeling will be conducted and reported elsewhere.

\section{Conclusion}
\label{sec:conclusion}

Through the calibration campaigns, we obtained the full energy response, temperature and bias dependence, and the full-energy peak effective area of the GRID detector, GRID-02. The energy response, including the channel-energy relation and energy resolution, can be used to generate the response matrix of the GRID detector. On the other hand, the effective area data helps validate the simulation model of the detector. The effective area at low energies confirms the detector’s on-axis absolute efficiency at energies near the gadolinium K-edge, while the internal mass distribution is verified based on the angular response, both with an overall discrepancy of less than 6\%. This validated model can be used for simulations, and the full response matrix will be created by adding the energy response to the model. \textcolor{red}{With a fit residual below 2.5\% for channel-energy relation and below 15\% for energy resolution, the mission requirements for the energy knowledge accuracy (less than 5\% error) and energy resolution accuracy (less than 15\% error) are both met. Thus, the generated response matrix will be a good estimation of the detector's spectral properties in data processing.}

After downloading the in-orbit observation data from the CubeSat, the data are first pre-processed using temperature and bias correction. The corrected data are then spectrally and temporally reconstructed using the response matrix. Astrophysical and other scientific studies can then commence with the reconstructed science data. Owing to the good characterisation of the detector during the calibration campaign, we discovered the possibility \textcolor{red}{of change} in the detector gain during long-term observations. This requires regular in-orbit calibration of the detector gain, using some known sources (e.g. $^{241}{\rm Am}$ onboard the CubeSat and cosmic annihilation radiation). By correcting the channel-energy relation, the influence of the change in the detector gain on the data integrity can be eliminated. \textcolor{red}{Since the SiPM bias is kept near the standard value (28.5 V) during normal operation, with the operating temperature ranging from 0~$^{\circ}{\rm C}$ to 20~$^{\circ}{\rm C}$, the residual error after the correction can be less than 3\%.}

The calibration of GRID-02 was performed using limited instruments for the measurements. The results are sufficient for in-orbit data process, but rough measurements lead to large errors and precision issues. The empirical function for the temperature and bias dependence is also based on a series of approximations and may not be a good representation of the actual detector response. Therefore, for the upcoming GRID detectors, \textcolor{red}{the} ground-based calibration will be performed \textcolor{red}{in} a setup similar to that of GRID-02, which will be gradually developed into a standard calibration procedure. To solve the problems of large errors in the results, improvements will be made to the measurement systems, for example, temperature monitors with better precision and better temperature controls. Further theoretical analysis of the detector, including its temperature and bias dependence and inner mass distribution, is also required. With perfected calibration methods and instrumentation, simulation models, and theoretical analysis, the production of scientifically valuable observation data for future GRID detector network will be feasible.

\begin{acknowledgements}
  This work is supported by Tsinghua University Initiative Scientific Research Program.
\end{acknowledgements}

\bibliographystyle{spbasic}
\bibliography{references.bib}

\end{document}